\documentstyle[epsf]{elsart}
\begin{document}

\def\note#1{{\bf[#1]}}

\def\dirac{{\bf \rm D}\!\!\!\!/\,}
\def\ddirac{{\bf \rm D}\!\!\!\!/\,_5}
\def\wilson{{\bf \rm W}}
\def\ham{{\bf \rm H}}
\def\mbham{{\cal H}}
\def\bmat{{\bf \rm B}}
\def\cmat{{\bf \rm C}}
\def\cond{{\rho_h}}

%\ltapprox and \gtapprox produce > and < signs with twiddle underneath
\def\spose#1{\hbox to 0pt{#1\hss}}
\def\ltapprox{\mathrel{\spose{\lower 3pt\hbox{$\mathchar"218$}}
 \raise 2.0pt\hbox{$\mathchar"13C$}}}
\def\gtapprox{\mathrel{\spose{\lower 3pt\hbox{$\mathchar"218$}}
 \raise 2.0pt\hbox{$\mathchar"13E$}}}
\def\inapprox{\mathrel{\spose{\lower 3pt\hbox{$\mathchar"218$}}
 \raise 2.0pt\hbox{$\mathchar"232$}}}

\begin{frontmatter}

\begin{flushright}
{\normalsize FSU-SCRI-98-71}\\
{\normalsize hep-lat/9807017}\\
\end{flushright}

\title{ 
A study of practical implementations of the Overlap-Dirac operator
in four dimensions}
\author{
Robert G. Edwards, Urs M. Heller and Rajamani Narayanan}
\address{
SCRI, The Florida State University, 
Tallahassee, FL 32306-4130, USA}

\begin{abstract}

We study three practical implementations of the Overlap-Dirac
operator $\dirac_o={1\over 2} [1 + \gamma_5\epsilon(\ham_w) ]$ 
in four dimensions.  Two implementations are based on different
representations of $\epsilon(\ham_w)$ as a sum over poles. One of them
is a polar decomposition and the other is an optimal fit to a
ratio of polynomials. The third one is obtained by representing
$\epsilon(\ham_w)$ using Gegenbauer polynomials and is referred to as
the fractional inverse method.  After presenting some spectral
properties of the Hermitian operator $\ham_o=\gamma_5\dirac_o$, we
study its spectrum in a smooth SU(2) instanton background with the aim
of comparing the three implementations of $\dirac_o$. We also present
some results in SU(2) gauge field backgrounds generated at $\beta=2.5$
on an $8^4$ lattice. Chiral properties have been numerically verified.

\end{abstract}

\end{frontmatter}

{\bf PACS \#:}  11.15.Ha, 11.30.Rd, 11.30.Fs\hfill\break

\section{Introduction}

The overlap formalism provides a way of realizing exact chiral
symmetry on the lattice. For vector gauge theories, there is now
promise that this formalism can be made practical.
The Overlap-Dirac operator derived from the overlap
formalism~\cite{overlap} of chiral fermions is~\cite{herbert1}
\begin{equation}
\dirac_o={1\over 2}[1 + \gamma_5\epsilon(\ham_w)]; \ \ \ \ \ 
\epsilon(\ham_w)= {\ham_w\over |\ham_w|}
\label{eq:ov_dirac}
\end{equation}
where 
\begin{eqnarray}
\ham_w &=& \pmatrix{\bmat - m & \cmat \cr \cmat^\dagger & -\bmat + m}; \\
\label{eq:hamil} 
\cmat_{i\alpha,j\beta}(n,n^\prime) &=& {1\over 2}
\sum_\mu \sigma_\mu^{\alpha\beta} \bigl[
U^{ij}_\mu(n)\delta_{n^\prime,n+\hat\mu} - (U^\dagger_\mu)^{ij}(n^\prime)
\delta_{n,n^\prime+\hat\mu} 
\bigr] \\
\label{eq:cmat}
\bmat_{i\alpha,j\beta}(n,n^\prime) &=& {1\over 2}\delta_{\alpha,\beta}
\sum_\mu \bigl[ 2\delta_{ij}\delta_{nn^\prime}-
U^{ij}_\mu(n)\delta_{n^\prime,n+\hat\mu} - (U^\dagger_\mu)^{ij}(n^\prime)
\delta_{n,n^\prime+\hat\mu}\bigr]
\quad
\label{eq:bmat}
\end{eqnarray}
is the Hermitian Wilson-Dirac operator. We will refer to $m$ as the
overlap mass. This is a parameter that has to be in the range $(0,2)$
for $\dirac_o$ to describe a single massless Dirac fermion.  In
principle, any value of $m$ in this region will yield the same
continuum theory. But at finite lattice spacing the cutoff effects can
be quite different, for different choices of $m$. In particular one
needs $m > m_1(g^2)$, for some $m_1(g^2)$ going to zero in the
continuum limit, in order for the overlap fermions to ``feel the
topology of background gauge fields''\cite{rough}.

It is necessary to efficiently deal with the action of
$\epsilon(\ham_w)$ on a vector for practical implementations of this
operator in four dimensional theories. Recently, one representation of
$\epsilon(x)$ as a sum of poles of $x^2$ was used to demonstrate that
a practical implementation of the Overlap-Dirac operator is possible
in three dimensions~\cite{herbert}.  One can also represent
$\epsilon(x)$ as an optimal rational function~\cite{Remez}. Another approach is
to use Gegenbauer polynomials to represent ${1\over \sqrt{x^2}}$
resulting in an iterative procedure to compute the action of
$\epsilon(\ham_w)$ on a vector~\cite{Bunk}.  In this paper we compare
these three methods from a practical point of view.  We start by
presenting some spectral properties of $\ham_o = \gamma_5\dirac_o$ and
$\dirac_o$ and then use a smooth SU(2) instanton background as an
example to illustrate these spectral properties. The example will also
enable us to compare the three practical implementations of
$\epsilon(\ham_w)$. We also present some results in SU(2) gauge field
backgrounds at $\beta=2.5$ on an $8^4$ lattice.

\section{Spectral properties of $\dirac_o$}

In this section, we build on some results of Neuberger~\cite{herbert1,herbert2} by
considering the eigenvalue problem of the Hermitian Overlap-Dirac operator,
\begin{equation}
\ham_o = \gamma_5\dirac_o = {1\over 2}[\gamma_5 + \epsilon(\ham_w)]\quad .
\label{eq:ov_ham}
\end{equation}
We do not use the explicit expression for $\ham_w$ in this section and
we only use the fact that $\epsilon^2(\ham_w)=1$. The exact chiral
symmetry built into the overlap formalism has been reemphasized in
the context of the fermionic action with $\dirac_o$ in Ref.~\cite{Luscher}. 
We therefore expect the spectrum of $\ham_o$ to essentially reproduce
all the important continuum properties.  The spectral properties of
the continuum operator $\ham=\gamma_5\dirac$ are
\begin{itemize}

\item Exact zero eigenvalues of $\ham$ are associated with topology.
The associated eigenvectors are also eigenvectors of $\gamma_5$. 
The zero eigenvalues are not paired in any sense. We can have $n_+$ zero
eigenvalues with positive chirality and $n_-$ zero eigenvalues with
negative chirality. The difference $(n_+-n_-)$ is the topology of
the background gauge field.

\item Non-zero eigenvalues of $\ham$ come in pairs that are equal
in magnitude and opposite in sign. The associated eigenvectors are
not eigenvectors of $\gamma_5$, but rather $\gamma_5$ has zero expectation
value in the eigenvectors, $\psi^\dagger \gamma_5 \psi = 0$.
\end{itemize}

On a finite four dimensional lattice, the matrix $\ham_o$ is a finite even 
dimensional matrix and $\det\ham_o$ is the overlap formula in a fixed
gauge background. The overlap exactly vanishes if the topology is non-trivial;
therefore, $\ham_o$ should have exact zero modes in a gauge field background
that is topologically non-trivial. These zero modes should have definite
chirality and the difference in the positive and negative modes is the
topology of the gauge field. If the topology is odd, then $\ham_o$ should
have an odd number of exact zero modes. Therefore it should also have
an odd number of non-zero modes. This implies that the number of positive
eigenvalues cannot be equal to the number of negative eigenvalues.
It is then interesting to explore the properties of the spectrum of
$\ham_o$ with the aim of establishing the equivalence of the two 
continuum properties listed above. We prove the following properties
of $\ham_o$ on the lattice:
\begin{itemize}

\item The spectrum of $\ham_o$ is bounded by the region $[-1,1]$.

\item Zero eigenvalues of $\ham_o$ have a definite chirality and need
not occur in pairs.  Eigenvalues of $\ham_o$ equal to $\pm 1$ also
have definite chirality.  These eigenvalues also need not occur in
pairs.

\item Non-zero eigenvalues of $\ham_o$ that have a magnitude less than
one come in pairs that are equal in magnitude and opposite in sign,
namely $\pm\lambda$.  The associated eigenvectors are not eigenvectors
of $\gamma_5$, but rather $\gamma_5$ has expectation value
$\pm\lambda$ in the eigenvectors, 
$\psi^\dagger \gamma_5 \psi = \pm\lambda$ for $\ham_o \psi = \pm\lambda \psi$.

\end{itemize}
Since $\ham_o$ is an even dimensional matrix, the unpaired zero eigenvalues
have to be matched by unpaired eigenvalues equal to $\pm 1$. This is what
is expected to happen in a topologically non-trivial background.

We present the proofs to the properties of $\ham_o$ listed above.
\begin{itemize}

\item
Squaring $\ham_o$ in (\ref{eq:ov_ham}) we get
\begin{equation}
\ham_o^2 = {1\over 4} \Bigl[ 2 + \gamma_5\epsilon(\ham_w) + 
\epsilon(\ham_w)\gamma_5 \Bigr ]
\label{eq:ov_hamsq}
\end{equation}
The above equation can be rewritten using (\ref{eq:ov_ham}) as
\begin{equation}
2\ham_o^2 = \gamma_5\ham_o + \ham_o\gamma_5
\label{eq:gin_wil}
\end{equation}
and is commonly referred to as the Ginsparg-Wilson relation.
As pointed out in~\cite{herbert1}, $\gamma_5\epsilon(\ham_w)$ is an
unitary operator and $\epsilon(\ham_w)\gamma_5$ is its Hermitian
conjugate. Therefore, all the eigenvalues of $[\gamma_5\epsilon(\ham_w) + 
\epsilon(\ham_w)\gamma_5]$ are bounded by $[-2,2]$. Therefore all
eigenvalues of $\ham_o^2$ are bounded by $[0,1]$ and from this the
first property of $\ham_o$ follows.

\item
Let us first find all eigenvectors of $\ham_o$ that are also
eigenvectors of $\gamma_5$. Let $\psi$ be an eigenvector of $\ham_o$
with eigenvalue $\lambda$ and let $\gamma_5\psi=\pm\psi$.  With the
aid of (\ref{eq:gin_wil}) one concludes that $\lambda(\lambda \mp 1)=0$. 
Therefore we have either $\lambda=0$ or $\lambda=\pm 1$, and we
have shown that chiral eigenvectors of $\ham_o$ have eigenvalues equal
to zero or $\pm 1$.  The subspace spanned by the zero eigenvalues of
$\ham_o$ can be split into two further subspaces. One with dimension
of $n_+$ will have positive chirality and one with dimension of $n_-$
will have negative chirality. The number $n_+$ need not be equal to
$n_-$ and if $(n_+ - n_-)$ is not equal to zero then the gauge field
carries a non-trivial topology. All eigenvalues of $\ham_o$ equal to
$+1$ have positive chirality and vice-versa.

\item 
Let $0 < \lambda < 1$ be an eigenvalue of $\ham_o$ with eigenvector
$\psi$ that is normalized to unity. Clearly, 
\begin{equation}
\phi = {\gamma_5\psi - (\psi^\dagger\gamma_5\psi)\psi \over
\sqrt{1-(\psi^\dagger\gamma_5\psi)^2} }; \ \ \ \ 
\psi^\dagger \phi = 0
\label{eq:phi}
\end{equation}
is a new vector that is orthonormal to $\psi$ since 
$\psi^\dagger\gamma_5\psi\ne \pm 1$. 
We can now compute $\ham_o\phi$ using the Ginsparg-Wilson relation,
(\ref{eq:gin_wil}), and find that
\begin{equation}
\ham_o\phi = -\lambda \phi + {2\lambda(\lambda-\psi^\dagger\gamma_5\psi) \quad .
\over \sqrt{1-(\psi^\dagger\gamma_5\psi)^2} } \psi
\label{eq:ham_phi1}
\end{equation}
This shows that $\ham_o$ acting on the subspace spanned by $(\psi,\phi)$
results in vectors in that subspace. From $\ham_o\psi=\lambda\psi$
and $\phi^\dagger\psi=0$, it follows that $\phi^\dagger\ham_o\psi=0$.
From the hermiticity of $\ham_o$, we also have $\psi^\dagger\ham_o\phi=0$.
This along with (\ref{eq:ham_phi1}) results in
\begin{equation}
\psi^\dagger\gamma_5\psi = \lambda
\label{eq:chiral1}
\end{equation}
since we have assumed $\lambda\ne 0$. We have also assumed that
$\lambda \ne 1$ in the beginning and therefore $\phi$ is indeed a
non-zero vector with
\begin{equation}
\ham_o\phi = - \lambda \phi;\ \ \ 
\phi = {\gamma_5\psi - \lambda \psi \over \sqrt{1-\lambda^2} } \quad .
\label{eq:ham_phi2}
\end{equation}
This establishes the last property of $\ham_o$. 

\end{itemize}

Comparing the properties of $\ham_o$ with the desired properties of
$\ham$ in the continuum, we conclude that $\ham_o$ reproduces topology
correctly, as expected. The eigenvectors of the non-zero eigenvalues
of $\ham_o$ do not have a zero expectation value of $\gamma_5$.
Eigenvalues of $\ham_o$ have to be scaled by the lattice spacing to
get the eigenvalues of the continuum operator and then only the
eigenvalues of $\ham_o$ close to zero have a continuum relevance.  The
last property tells us that the corresponding eigenvectors also have
an expectation value of $\gamma_5$ close to zero as expected in the
continuum.

It is trivial to find the eigenvalues and left eigenvectors of
$\dirac_o$ since $\dirac_o$ also maps the subspace 
spanned by $(\psi,\phi)$ onto itself. The eigenvalues of
$\dirac_o$ are zero when $\lambda=0$, unity when $\lambda=\pm 1$
and the complex conjugate pairs $(\lambda^2 \pm i\lambda\sqrt{1-\lambda^2})$
for $0 < \lambda < 1$. This proves that the only real eigenvalues of
the Overlap-Dirac operator are zero and unity. The massive 
Overlap-Dirac operator is given by $\dirac_o + m_f$ with
$-1/2 < m_f < \infty$~\cite{herbert3}~\footnote{
The fermion mass $m_f$ should be written as $m_f={\mu\over 1-\mu}$ with
$\mu$ in the range $[-1,1]$. Positive values of $\mu$ correspond to
positive quark masses and vice-versa.~\cite{herbert3}}.
Clearly the operator has a
zero eigenvalue only when $m_f=0$. Occurrence of zero eigenvalues
for a lattice Dirac operator at positive quark masses in certain
gauge field backgrounds are referred to as ``exceptional'' configurations.
Such ``exceptional'' configurations are actually generic for the Wilson-Dirac
operator but non-existent for the Overlap-Dirac operator. As pointed
out in Ref.~\cite{herbert4} the only difficulty arises in defining the
Overlap-Dirac operator if $\ham_w$ has a zero eigenvalue since
$\epsilon(\ham_w)$ is then not well defined. In such a background configuration,
one can define the operator at a value of $m$ slightly away from the
one where $\ham_w$ has a zero eigenvalue since the continuum limit is
expected to be independent of $m$.

Finally, one can also trivially get the spectral flow of the massive
$\ham_o(m_f)$.  If $\ham_o(0)$ has a zero eigenvalue then
$\ham_o(m_f)$ has an eigenvalue equal to $\pm m_f$ corresponding to
positive and negative chirality of the eigenvector.  If $\ham_o(0)$
has an eigenvalue equal to $\pm 1$, then $\ham_o(m_f)$ has an
eigenvalue equal to $\pm (1+m_f)$.  For the paired eigenvalues $\pm
\lambda$ of $\ham_o(0)$ in the range $0 < \lambda < 1$, $\ham_o(m_f)$
has paired eigenvalues equal to 
$\pm\sqrt{ \lambda^2 + 2\lambda^2m_f + m_f^2 }$. This again shows that
zero crossings in the region $ -1/2 < m_f < \infty$ occur only at
$m_f=0$ and these are due to the topological content of the gauge
field.

\section {Algorithms}

The key element in the application of the Overlap-Dirac operator
(\ref{eq:ov_dirac}) on a vector is how to apply $\epsilon(\ham_w)$ on a
vector. The meaning of $\epsilon(\ham_w)$ is defined through the
$\epsilon$ function on the eigenvalues of $\ham_w$. Hence, the
diagonalization of $\ham_w$ is one possible method and has been
applied in two dimensional systems~\cite{schwinger}. 
However, on larger systems of interest, exact diagonalization is
prohibitive.

\subsection{Fractional inverse method}

By representing $\epsilon(\ham_w)$ as 
\begin{equation}
\epsilon(\ham_w) = \left(\ham_w^2\right)^{-1/2} H_w
\label{eq:eps_frac_inv}
\end{equation}
we can directly apply $\epsilon(\ham_w)$ on a vector by using a Krylov
space iterative solver for the fractional inverse based on Gegenbauer
polynomials \cite{Bunk}. In this method, one parameterizes
the system to be solved,
\begin{equation}
M^\gamma \phi = \chi\quad ,
\label{eq:frac_eq}
\end{equation}
by
\begin{equation}
M = c ( 1 + t^2 - 2 t A)\quad ,
\label{eq:gegen_param}
\end{equation}
The key to solving (\ref{eq:frac_eq}) is then the generating function for
Gegenbauer polynomials
\begin{equation}
( 1 + t^2 - 2 t A)^{-\gamma}  = \sum_{n=0}^{\infty} t^n C_n^\gamma(A)\quad .
\label{eq:gegen_gen}
\end{equation}
Using the recursion relations for the $C_n^\gamma$
one can construct an iterative solver for $\phi$ in (\ref{eq:frac_eq}).
We refer the reader to \cite{Bunk} for details. 
For the case of interest, $\gamma = 1/2$ and $M=\ham_w^2$ is
hermitian and positive (semi-)definite. 
We restrict the bounds of $A$ to be in $[-1,1]$ and obtain uniform
convergence of $A$ and hence $M = \ham_w^2$ in (\ref{eq:gegen_gen})
provided  $|t| < 1$. It is important for the best convergence to match
the extremal eigenvalues of 
$\ham_w^2 \in [\lambda_{min}^2,\lambda_{max}^2]$ to those of $A \in
[-1,1]$. Then, the $t$ and $c$ are determined and
%
% t = {{\sqrt(\kappa) - 1} \over {\sqrt(\kappa) + 1}}
%
\begin{equation}
t = 1 - {1 \over \kappa} ,
\label{eq:conv_fact}
\end{equation}
where $\kappa = |\lambda_{max} / \lambda_{min}|$ is the condition
number of $\ham_w$. The parameter $t$ is the convergence factor for
the algorithm.  Typically, $\lambda_{max} \ltapprox 8$, but
$|\lambda_{min}|$ can be quite small \cite{rough}. Control of the
lowest mode is important since the number of iterations expected to
reach a given accuracy $R$ is bounded by $-\log(R)\kappa$, and
numerical tests have shown this bound to be saturated. We remark that
the same bound also applies to the Conjugate Gradient (CG) algorithm
but is not expected to be saturated. The polynomials generated by the
Gegenbauer expansion are not optimal, while the CG algorithm is
optimal.  Eigenvectors with small eigenvalue can be projected out of
(\ref{eq:eps_frac_inv}) to improve convergence.

\subsection{Pole methods}
\subsubsection{Polar decomposition}\label{sec:tanh}

Another approach is to consider $\epsilon(z)$ as the limit of some function
that switches quickly between $\pm 1$ on $z$ changing sign.
Recently, Neuberger showed \cite{herbert,pensacola} that 
\begin{equation}
\epsilon(z) = \lim_{N\rightarrow\infty} f_N(z) ,
\end{equation}
with
\begin{equation}
f_N(z) = {{(1+z)^{2N} - (1-z)^{2N}} \over {(1+z)^{2N} + (1-z)^{2N}}} \quad .
\label{eq:f_N}
\end{equation}
In particular, $f_N(z)$ provides a good approximation for
$\epsilon(z)$ at finite $N$ for $1/N \ltapprox z \ltapprox N$, and can
be expressed quite simply, by matching poles and residues, as a sum over
poles
\begin{equation}
f_N(z) = z \sum_{k=1}^N  {{a_k} \over {z^2 + b_k}}\nonumber
\end{equation}
\begin{equation}
a_k = {1 \over {N \cos^2({\pi\over{2 N}}(k - \frac{1}{2}))}}\ ,\qquad
b_k = \tan^2({\pi\over{2 N}}(k - \frac{1}{2}))\quad .
\label{eq:tanh_pole}
\end{equation}
Since $\epsilon(s z) = \epsilon(z)$ for positive $s$, the argument $z$
can be rescaled into the region suitable for the approximation. 
% This is especially important for the lowest modes.
For the application of $f_N(\ham_w)$ on a vector, one can use a
multi--shift CG solver \cite{multi_shift} for the $N$ terms in
$f_N(\ham_w)$ \cite{herbert}. The convergence is governed by the
smallest shift; however, the overall cost is not much more than the
inversion for the smallest shift. The representation of $f_N(\ham_w)$
in terms of poles is the key observation to making the algorithm
practical.

While the smallest shift, $b_1$, in (\ref{eq:tanh_pole}) governs
convergence, the choice of $N$ is determined by the condition number
of $\ham_w$.  The optimal choice for the scale factor $s$ is 
${1\over \sqrt{\lambda_{max} \lambda_{min}}}$. The maximum deviation
of $f_N(sz)$ away from unity occurs at $z=\lambda_{max}$ and
$z=\lambda_{min}$, and the maximum deviation monotonically decreases
in $N$.  The desired accuracy $\delta$ over the spectral range of
$\ham_w$ fixes $N$ depending on the condition number $\kappa$, namely
$N=0.25\sqrt\kappa\ln{\delta\over 2}$.  Hence, the condition number is
the real measure of convergence for small $\lambda_{min}$ if a good
uniform approximation is to be achieved.  This requirement may not be
needed in practice as long as the lowest modes are well represented,
but will be used throughout this work since it is not difficult to
achieve.

% It is amusing to note that $f_N(z)$ of (\ref{eq:f_N}) is a very good
% approximation to $\tanh(2 N z)$, which, of course, in the limit $N \to \infty$
% rapidly approaches $\epsilon(z)$. This leads to a connection with the
% domain wall approach, where the determinant of the five dimensional Dirac
% operator, after dividing by the contribution from the pseudofermion fields is
% \cite{herbert3}
% \note{MAKE 1/2 FACTORS AGREE WITH LATER TANH}
% %
% \begin{equation}
% {\rm det}(\ddirac) = {\rm det}\left({1\over 2}\left[1 +
%    \gamma_5 \tanh(\frac{1}{2} L_s \ham_d) \right]\right)
% \label{eq:domain_wall}
% \end{equation}
% %
% where $\ham_d$ is the log of the five dimensional domain wall transfer
% matrix, and $L_s$ is the extent of the extra dimension. In the
% $L_s \rightarrow\infty$ limit, $\tanh(L_s z)\rightarrow \epsilon(z)$.
% Going to the continuous hamiltonian limit $\ham_d\rightarrow\ham_w$,
% we recover the determinant of the Overlap-Dirac operator in
% (\ref{eq:ov_dirac}).

\subsubsection{Optimal rational approximation}

Other approximations to $\epsilon(z)$ over a finite interval can be
suggested. A polynomial approximation like Chebyshev over the interval
$[-1,1]$ is particularly bad since it is only logarithmically
convergent. A better choice is the ratio of polynomials called an
optimal rational function approximation.  Since $\epsilon(z)$ is odd, we
choose the form
\begin{equation}
g_N(z) = z {{P_N(z^2) \over {Q_N(z^2)}}}
\label{eq:ratio_approx}
\end{equation}
where we assume the polynomials are irreducible of degree $N$. We
choose the highest order coefficient in $Q_N(z)$ to be one. The
normalization is contained in $P_N(z)$ and is defined to be $c_0$.  
As in the previous section, since $\epsilon(s z) = \epsilon(z)$ for
positive $s$, the argument $z$ can be rescaled into the region
suitable for the approximation.

To cast into a suitable form, we change variables to $y = z^2$ and
consider $h(y) = (1/z) \epsilon(z)$ over the interval $[0,z_{max}^2]$.
One can show \cite{Rivlin} that there exist best approximations of
$r(y) = P_N(y)/Q_N(y)$ for a continuous function in a uniform norm
over the interval $[y_{min},1]$ (with no loss of generality, we
restrict the upper bound). The deviation of $r(y)$ from the intended
function $1/\sqrt{y}$ can be shown~\cite{Rivlin} to have $2 + 2 N$
extrema with two always the endpoints.
An algorithm, sometimes called the Remez algorithm
\cite{Remez,Rivlin,Moshier}, exists to find the best approximation
$r(y)$. The version used here requires the extrema of the relative
error to be constant. For our application, this implies the worst
error in $g_N(z)$ is for the smallest $z$. Other weighting functions
can be chosen.

The numerical evaluation of (\ref{eq:ratio_approx}) is most stable in
factored form
\begin{equation}
g_N(z) = z c_0\,{{\prod_{k=1}^N (z^2 + p_k)} \over 
  {\prod_{k=1}^N (z^2 + q_k)}}
\label{eq:ratio_factored}
\end{equation}
where we find the roots $p_k$ and $q_k$ to be real and positive.
We can cast this expression as a sum over poles as in Section~\ref{sec:tanh}:
\begin{equation}
g_N(z) = z \left(c_0 + {\sum_{k=1}^N {c_k \over {z^2 + q_k}}}\right);\qquad
c_k = c_0 {{\prod_{i=1}^N (q_k - p_i)} \over 
  {\prod_{i=1,i\ne k}^N (q_k - q_i)}}
\label{eq:ratio_pole}
\end{equation}

\begin{figure}
\epsfxsize=5in
\centerline{\epsffile{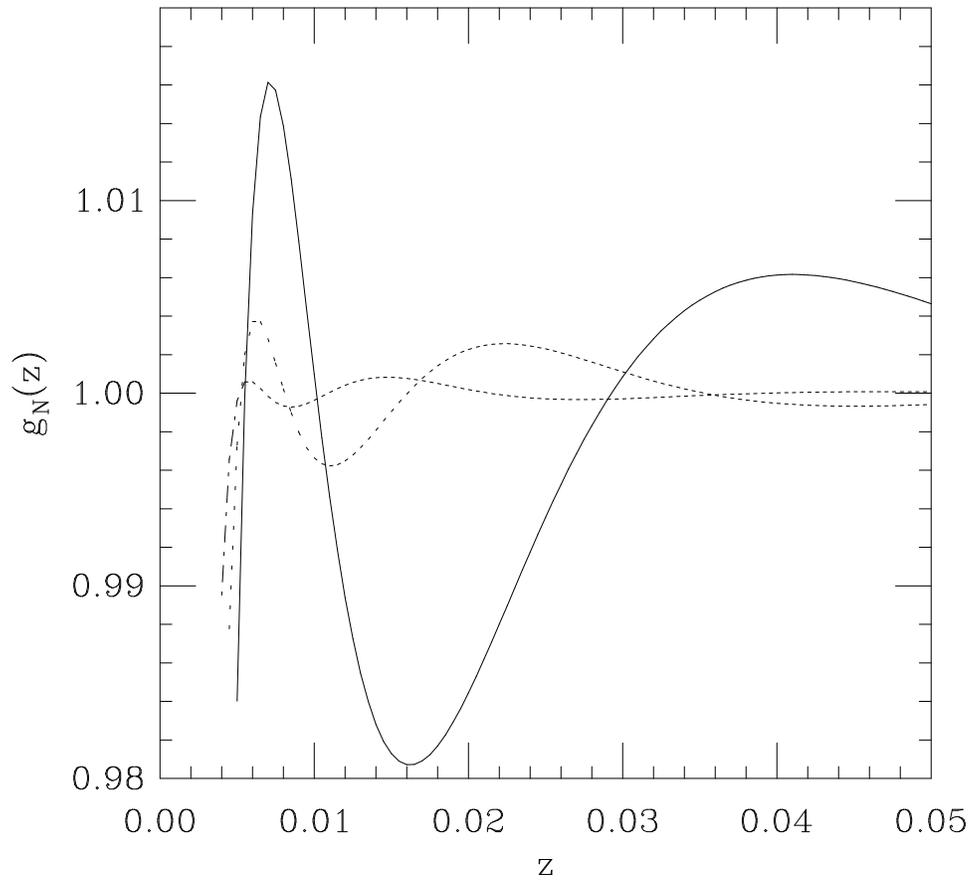}}
\caption{
The optimal rational function approximation $g_N(z)$ to $\epsilon(z)$ in the interval
$[0.005,1]$ for $N=6, 8$ and $10$. Shown is a detail of the region
of largest deviation. Outside this region, the curves rapidly approach
unity.
}
\label{fig:remez}
\end{figure}

We choose the value of $N$ so that the deviations of $|g_N(z)|$ over
the interval $[z_{min},1]$ are some acceptably small value
$\delta$. We show examples in Fig.~\ref{fig:remez} of $N=6, 8$ and
$10$ for the fit interval $[0.005,1]$, which admits a condition number
of $200$. We see the method rapidly converges in $N$. In our tests, we
have chosen $N=8$ where we see less than $0.5\%$ of an error. However,
we observe that since we are approximating a flat function, we can in
practice extend the acceptable upper bound beyond $1$. For $z > 1$,
$g_N(z)$ monotonically increases. For $N=8$, we can choose 
$z_{max} = 2.8$ and still maintain a $1\%$ error. We can use this trick to
effectively increase the useful allowable condition number to over
$500$. In Table~\ref{tab:remez}, we show the roots and normalizations
for the polynomials shown in Fig.~\ref{fig:remez}.

\section{Results}

We will consider a gauge field background with a
single SU(2) instanton to compare the performance of
the three algorithms described in the previous section.
The SU(2) link elements on the lattice obtained from 
an exact evaluation of the path ordered integral of the
continuum instanton take the following form~\cite{Laursen}. 
In the singular gauge,
\begin{equation}
U_\mu(n) = \exp \Bigl[ib_\mu(n)\cdot\sigma 
\bigr(\vartheta_\mu(n;0)-\vartheta_\mu(n;\rho)\bigl)\Bigr]
\label{eq:Inst_s}
\end{equation}
\begin{equation}
\vartheta_\mu(n;\rho) = {1\over \sqrt{ \rho^2 + \sum_{\nu\ne\mu} 
(n_\nu-c_\nu)^2}}
\tan^{-1} { \sqrt{ \rho^2 + \sum_{\nu\ne\mu} (n_\nu-c_\nu)^2} \over
\rho^2 + \sum_{\nu} (n_\nu-c_\nu)^2 + (n_\mu-c_\mu) }
\label{eq:phi_inst}
\end{equation}
\begin{eqnarray}
b_1(n) & = & (-n_4+c_4, n_3-c_3, -n_2+c_2),\cr
b_2(n) & = & (-n_3+c_3, -n_4+c_4, n_1-c_1),\cr
b_3(n) & = & (n_2-c_2, -n_1+c_1, -n_4+c_4),\cr
b_4(n) & = & (n_1-c_1, n_2-c_2, n_3-c_3)
\label{eq:b's}
\end{eqnarray}
Here $c$ denotes the center of the instanton, and $\rho$ is the size of
the instanton measured in lattice units. 
We work on an $8^4$ periodic lattice and set $\rho=1.5$ and
$c_\mu=4.5$ with the lattice sites in each direction numbered from $1$ to $8$.

\begin{figure}
\epsfxsize=5in
\centerline{\epsffile{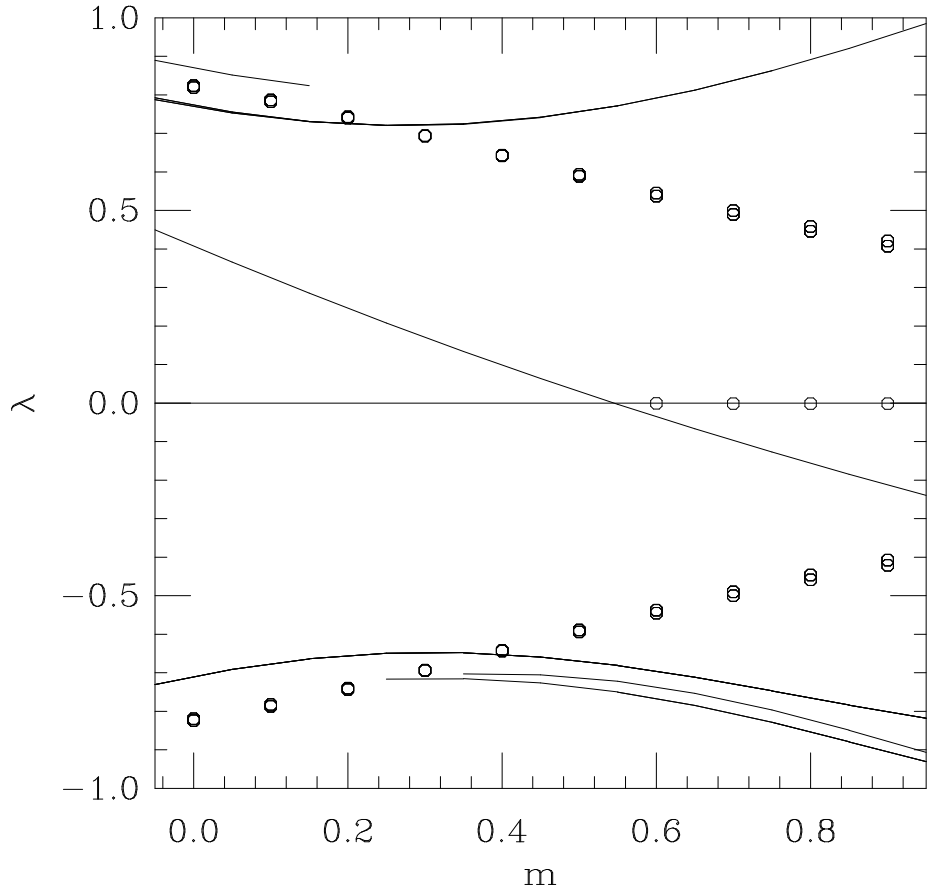}}
\caption{
The octagons describe the low lying spectrum of $\ham_o$ in
an smooth SU(2) instanton background.
The zero mode found at $m > 0.5$ is singly degenerate and is associated
with the instanton. 
The lines are the spectral flow of $\ham_w$.
}

\label{fig:flow_inst}
\end{figure}

\begin{figure}
\epsfxsize=5in
\centerline{\epsffile{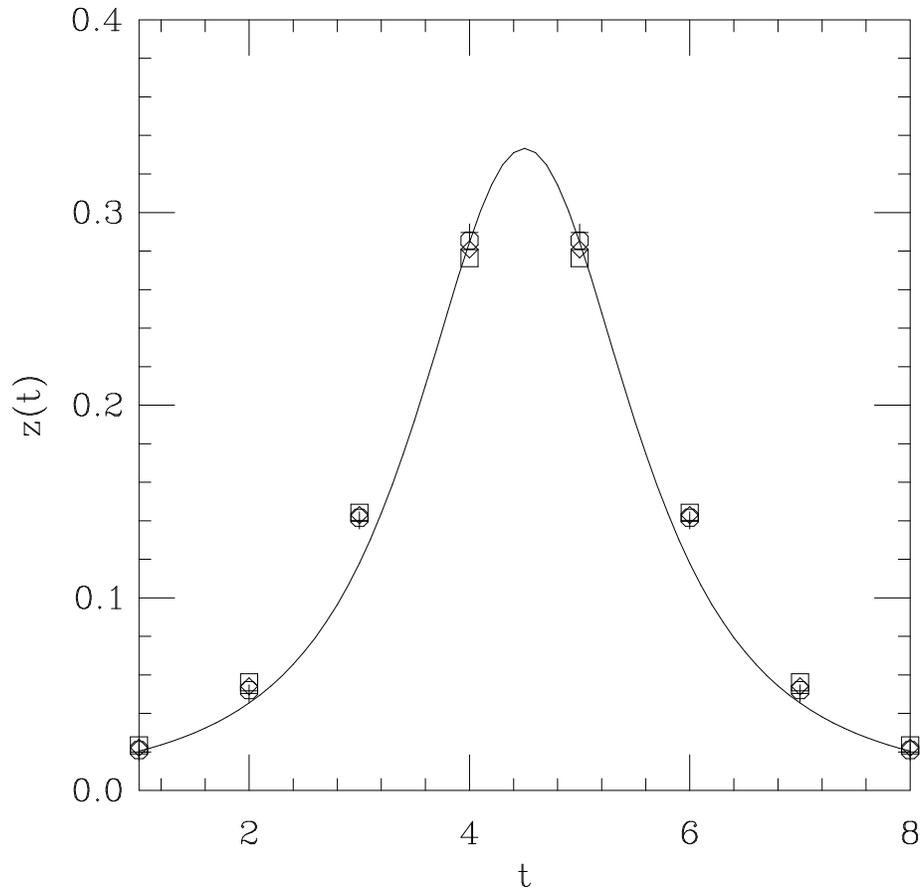}}
\caption{
The mode associated with the zero eigenvalues in
Fig.~\ref{fig:flow_inst} at $m=0.6$ (square), $m=0.7$ (diamond).
$m=0.8$ (octagon) and $m=0.9$ (plus).
The solid line is the continuum
zero mode
with $\rho=1.5$.
}
\label{fig:zero_inst}
\end{figure}

We first present the results. We used the explicitly real
representation of $\ham_w$~\cite{herbert2}.  We computed a few low
lying eigenvalues of $\ham_o$ using the Ritz algorithm~\cite{ritz} to
compute the lowest eigenvalues of $\ham^2_o$. From our previous work
on the spectral flow of the hermitian Wilson-Dirac operator, we know
that a level of $\ham_w$ crosses zero between $m=0.5$ and
$m=0.6$~\cite{smooth}. This flow is shown as a solid line in
Fig.~\ref{fig:flow_inst}. The low lying eigenvalues of $\ham_o$ as a
function of $m$ are also shown in the same figure as octagons.  For 
$0 \le m \le 0.5$, fourteen low lying eigenvalues are shown and for 
$0.6 \le m \le 0.9$ fifteen low lying eigenvalues are shown. The non-zero
eigenvalues come in opposite pairs with their chirality being equal to
their eigenvalue.  This is in accordance with the last property of
$\ham_o$.  The seven positive and the seven negative eigenvalues are nearly
degenerate.  For $0.6 \le m \le 0.9$, there is a single zero
eigenvalue for $\ham_o$. This is due to the instanton background.
The abrupt appearance of a zero eigenvalue with chirality $+1$ is
accompanied with the appearance of an eigenvalue of $-1$ with chirality
$-1$. It is difficult to confirm this numerically because of the
observed high density of eigenvalues near $\pm 1$.
The shape of the zero mode for $0.6 \le m \le 0.9$ is shown in
Fig.~\ref{fig:zero_inst}. 
The $z(t)$ plotted is equal to $\sum_{\vec x a} \psi^2(\vec x, t, a) $
and the continuum mode is plotted as a solid line. The zero mode is
remarkably stable under change of the overlap mass $m$.

\begin{figure}
\epsfxsize=5in
\centerline{\epsffile{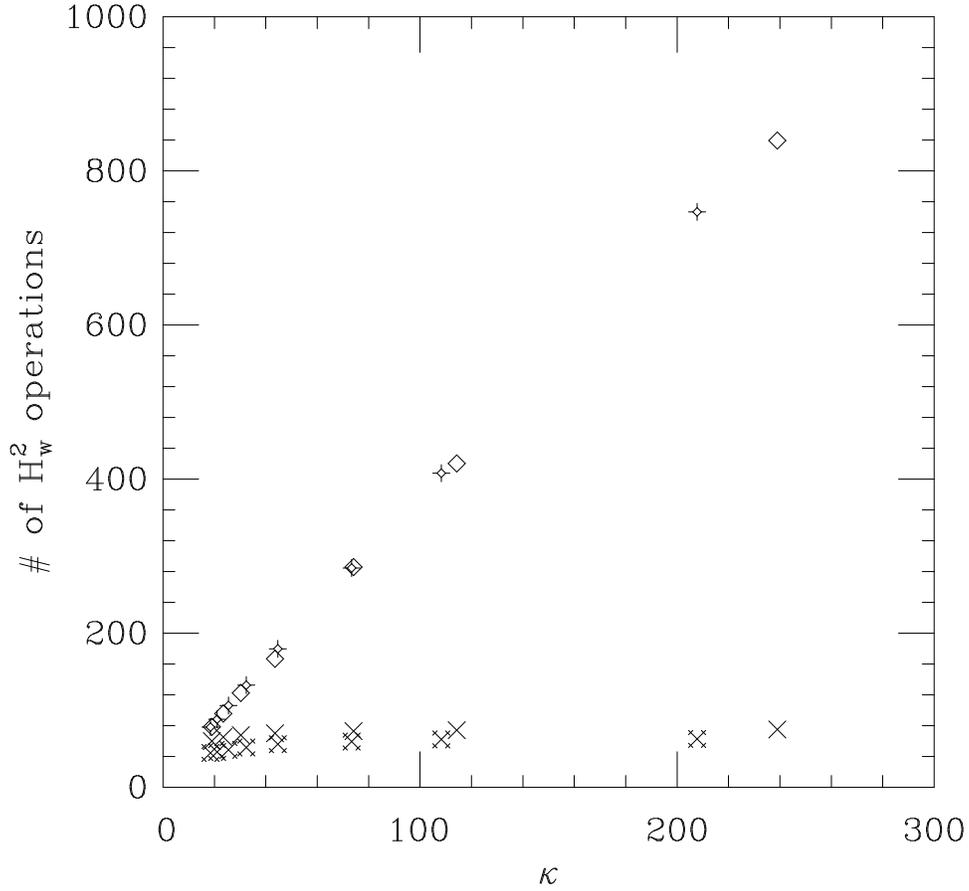}}
\caption{
Average number of $\ham_w^2$ operations for the optimal rational approximation
and the fractional inverse methods to apply $\epsilon(\ham_w)$ on a
vector versus the Wilson-Dirac operator condition number for the
configuration in Fig.~\ref{fig:flow_inst}. 
The cross and fancy cross are for the optimal rational approximation below and
above the crossing, resp. The diamond and fancy diamond are for the
fractional inverse method below and above the crossing, resp.
}
\label{fig:iters_inst}
\end{figure}

We have made comparisons of the average number of $\ham_w^2$
operations to obtain the result of the action of $\epsilon(\ham_w)$ on
a vector.  The average is over all the Ritz iterations needed in
solving for the lowest eigenvalue in Fig.~\ref{fig:flow_inst}. The
main comparison is between the pole method and the fractional inverse
method. The two pole methods are expected to be comparable in the
average number of $\ham_w^2$ operations on a vector. Since we are
using the method of pole shifts the number of poles only affect the
memory requirement and some additional multiplications of vectors by
scalars and additions of vectors.  We show in
Fig.~\ref{fig:iters_inst} the scaling of the number of $\ham_w^2$
operations versus the condition number $\kappa$ of $\ham_w$ at masses
below and above the crossing. We used the $N=8$ polynomial in
Tab.~\ref{tab:remez} (c.f. Eq.~(\ref{eq:ratio_pole})). The fractional
inverse method scales quite linearly in $\kappa$ and the $\ham_w^2$
operation count grows to large values while in the optimal rational
approximation it grows very slowly. This is because the bound on
the rate of convergence is saturated in the fractional inverse method
while in CG it is not.  As expected, no substantial difference in the
number of $\ham_w^2$ operations was found between a $N=50$ order
$f_N(\ham_w)$ in Eq.~(\ref{eq:tanh_pole}) and the $N=8$ optimal rational
approximation.

Although the two pole methods need similar number of $\ham_w^2$
operations there are some differences. For a fixed accuracy (maximum
deviation $\delta$ away from unity in the realization of
$\epsilon(z)$) and a fixed condition number the number of poles needed
in the optimal rational approximation is smaller than the one needed in the
polar decomposition. For example, with $\delta = 0.01$, the
optimal rational approximation needs seven poles for a condition number of 200
and twelve poles for a condition number of 1000. In contrast,
the polar decomposition needs 19 poles for a condition number of 200
and 42 poles for a condition number of 1000. This indicates that the
number of poles needed grows with the condition number at a slower
rate in the optimal rational approximation than in the polar decomposition.  

\begin{figure}
\epsfxsize=5in
\centerline{\epsffile{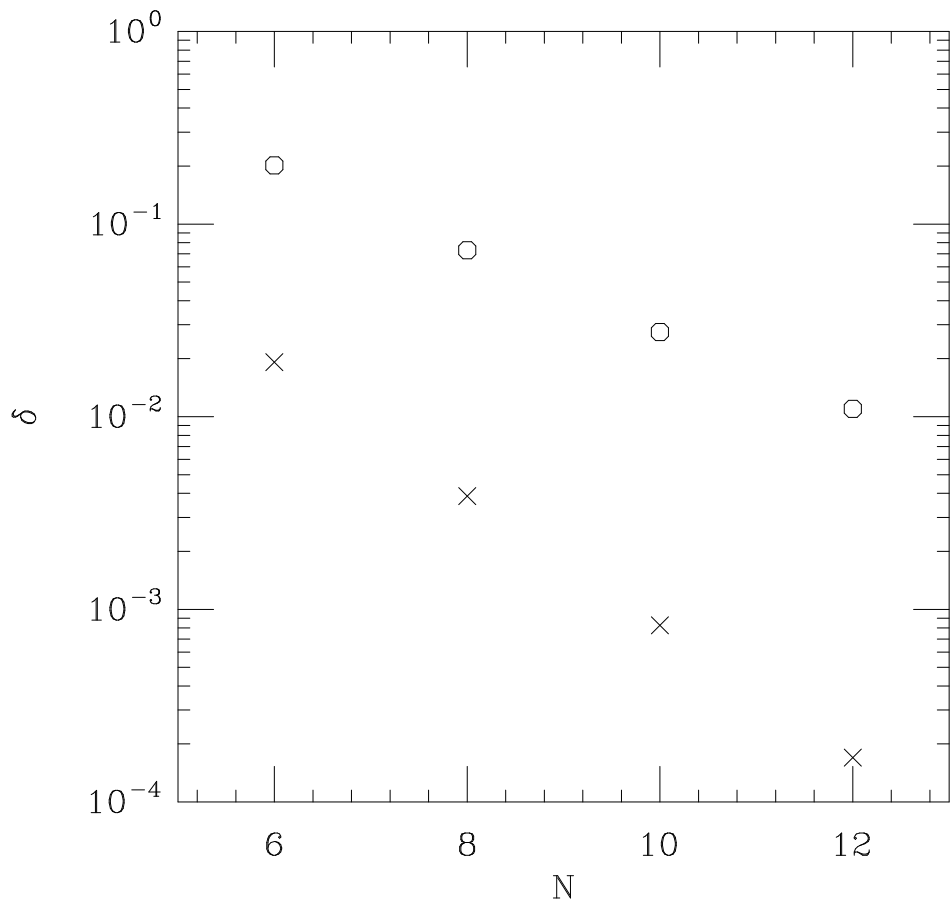}}
\caption{
The accuracy as a function of the number of poles
for a fixed condition number. The crosses are for condition number
$\kappa=200$ and the octagons are for $\kappa=1000$.
}
\label{fig:error}
\end{figure}

A comparison of the number of poles needed for the optimal rational
approximation as a function of the accuracy for a fixed condition
number is shown in Fig.~\ref{fig:error}.  The number of poles needed
depends logarithmically on the accuracy, $\delta$, and the slope is
smaller for larger condition number.  As mentioned before, the number
of poles needed only affects the memory requirement and results in a
few additional operations of two types: addition of two vectors and
multiplication of vector by a scalar.  For a fixed memory requirement,
the optimal rational approximation will admit a larger condition number.

\begin{figure}
\epsfxsize=5in
\centerline{\epsffile{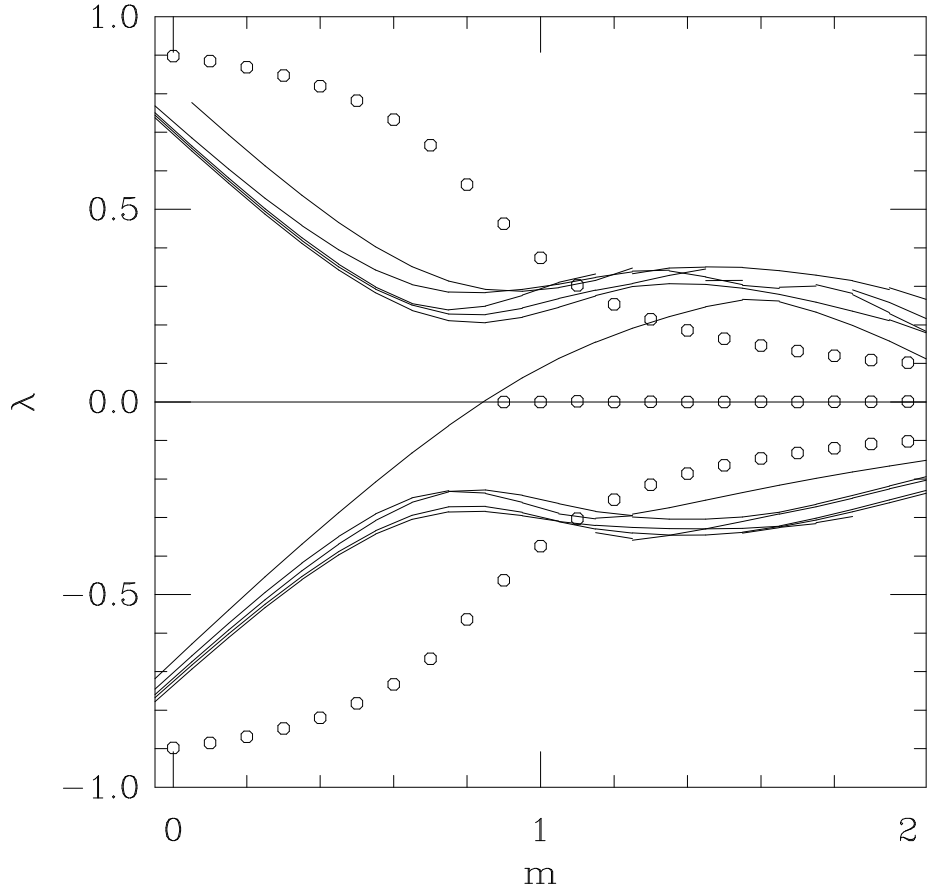}}
\caption{
The octagons describe the low lying spectrum of $\ham_o$
in a pure SU(2) gauge field background at $\beta=2.5$.
The zero mode found at $m > 0.8$ is singly degenerate and is associated
with an instanton in the background gauge field. The lines are the spectral flow
of $\ham_w$.
}
\label{fig:flow_cfg17}
\end{figure}

\begin{figure}
\epsfxsize=5in
\centerline{\epsffile{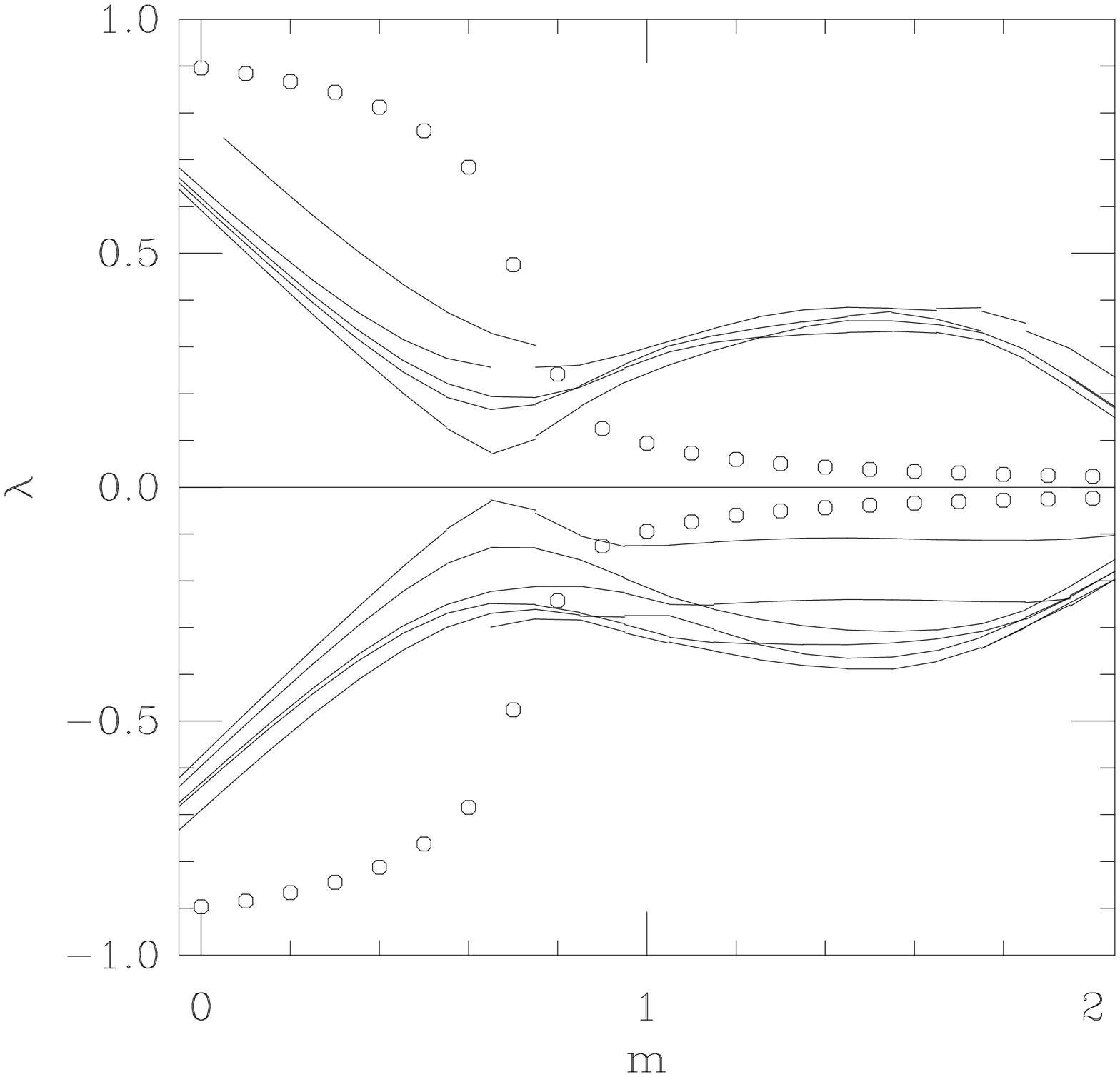}}
\caption{
Same as in Fig.~\ref{fig:flow_cfg17} for a different SU(2) configuration
at $\beta=2.5$.
}
\label{fig:flow_cfg1}
\end{figure}

\begin{figure}
\epsfxsize=5in
\centerline{\epsffile{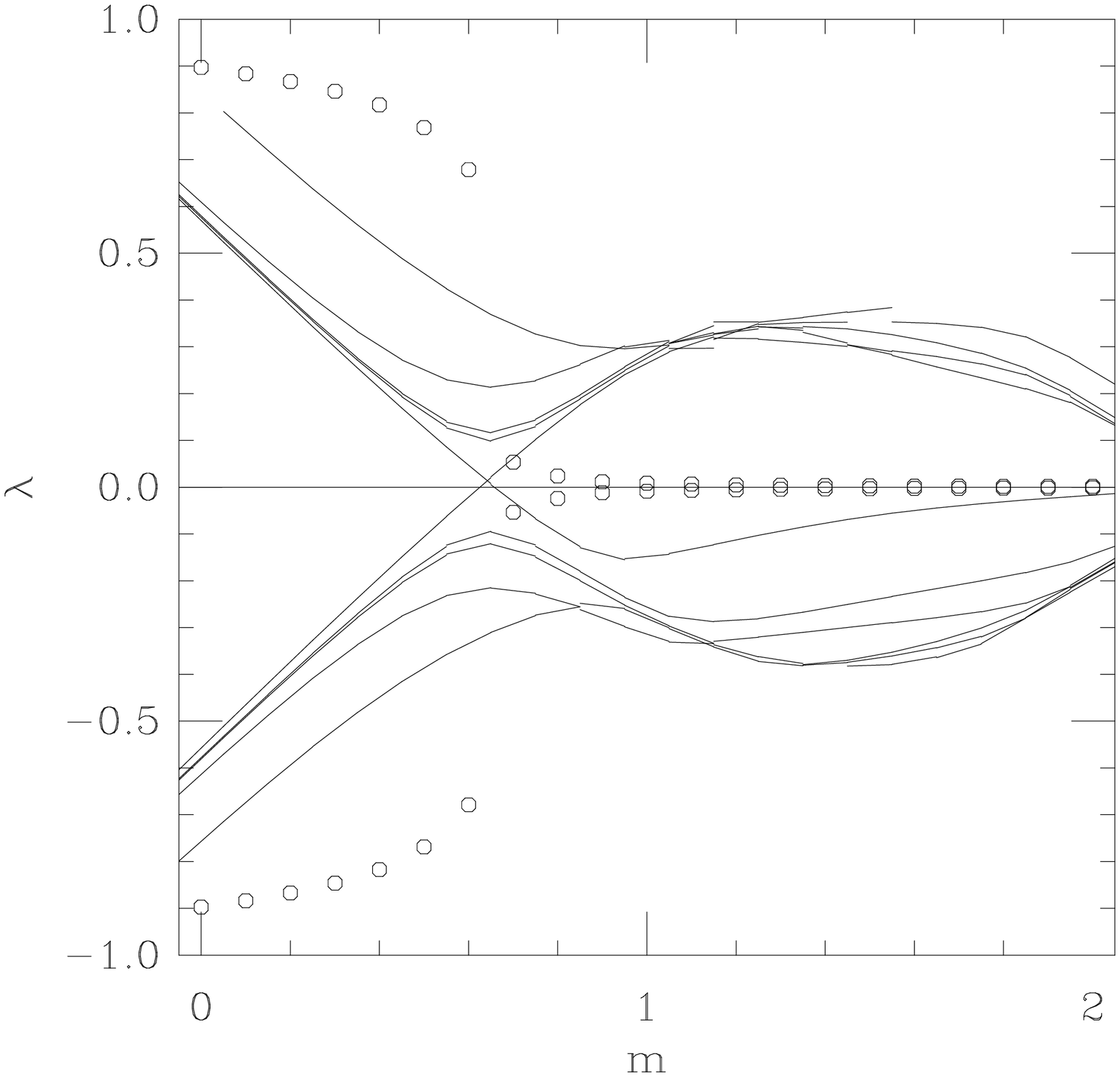}}
\caption{
Same as in Fig.~\ref{fig:flow_cfg17} for a different SU(2) configuration
at $\beta=2.5$.
}
\label{fig:flow_cfg11}
\end{figure}

We also present the results of the spectral flow of $\ham_o$ in
three pure SU(2) gauge field backgrounds generated at $\beta=2.5$
on an $8^4$ lattice to further emphasize the feasibility of
the algorithms described here. The flows are shown in 
Fig.~\ref{fig:flow_cfg17}--\ref{fig:flow_cfg11}.  
Fig.~\ref{fig:flow_cfg17} is the flow in
a background that contains a single instanton 
and this results in an exact zero mode of
$\ham_o$. We verified that it is an exact zero mode by computing
its chirality which was found to be equal to $-1$.
Fig.~\ref{fig:flow_cfg1} is the flow in
a background that has no topological object and we do not find
any exact zero mode in this case.
Fig.~\ref{fig:flow_cfg11} is the flow in
a background that contains an instanton and an anti-instanton.
These will result in approximate zero modes and one can see
evidence for this in the sharp drop in the flow between
$m=0.6$ and $m=0.7$ and also by comparing to the flow 
in Fig.~\ref{fig:flow_cfg1}. The modes in Fig.~\ref{fig:flow_cfg11}
are definitely not exact zero modes and, in accordance with the
last property of $\ham_o$, have chiralities equal to their eigenvalues.
Finally we also show
the exact zero mode found in the flow in Fig.~\ref{fig:flow_cfg17}.
We plot it for all $m >0.8$ and find that the mode is essentially
the same and it also agrees with the zero mode of $\ham_w$ 
computed close to the point in $m$ where $\ham_w$ has a zero crossing.

\section{Concluding remarks}

We have compared three implementations of the Overlap-Dirac operator
in four dimensions.  We find that the pole methods are superior in
convergence to the fractional inverse method.  The main advantage of
the optimal rational function approximation is that a suitable error can be
achieved for a smaller order polynomial.  This saves some work in the
multi-shift CG since fewer vectors are need. Also, memory can be
saved. No difference in convergence properties is expected otherwise.

For the massive Overlap-Dirac operator, one can again use the 
multi-shift CG solver~\cite{multi_shift} to solve for the inverse with
several quark masses. The action of $\epsilon(\ham_w)$ is only applied
on one vector and the quark masses can be expressed as a shift.

In practice it may be worth projecting out a few small eigenvectors of
$\ham_w$ before the action of $\epsilon(\ham_w)$ on a vector.  This is
useful if there are a few well separated small eigenvalues.  The
projection can have a dramatic effect in the fractional inverse method
since the number of $\ham_w^2$ operations is linear in the condition
number. It will also decrease the operation count in the pole methods.
We have verified these effects on the configurations studied in this
paper.

The pairing of the non-zero eigenvalues of $\ham_o$ can be used
to reduce the computation of the spectrum of $\ham_o$.

We have demonstrated the feasibility of using the Overlap-Dirac
operator in a four dimensional gauge theory by computing the spectral
flow of $\ham_o$ in SU(2) gauge field backgrounds at $\beta=2.5$ on an
$8^4$ lattice. Chiral properties have been shown to be properly
reproduced.  Exact zero modes have been shown to be related to
topology. They are clearly differentiated from almost zero modes in
that the exact zero modes have chirality $\pm 1$ while the almost zero
modes have chirality equal to their eigenvalue, namely very small.

% Results for each method after projecting out the smallest eigenvector
% in applying $\epsilon(\ham_w)$ are also shown in
% Fig.~\ref{fig:iters_inst}.  The projections saves enormously on
% iterations in the fractional inverse method since the condition number
% is effectively reduced. Scaling is still linear.  \note{CHECK ALL
% THIS}.  The number of iterations for the optimal rational approximation is
% halved.  In all cases, the decrease will be less if in a configuration
% the density of small eigenvalues is high.

% For domain wall fermions, there is no freedom of an additional scale
% factor $s$, and a small eigenvalue of $\ham_w$ necessitates a suitable
% large $L_s$ in (\ref{eq:domain_wall}) to avoid a strong $L_s$
% dependence in the determinant of the Dirac operator. In this latter
% regard, the domain wall and the approximations used for
% $\epsilon(\ham_w)$ in the overlap formalism are the same since the small
% eigenvalues of $\ham_w$ are the same as for $\ham_d$
% \cite{something_here}  \note{IMPROVE THIS}.

\ack{
The authors would like to thank Tony Kennedy and Stefan Sint for
useful discussions. This research was supported by DOE contracts
DE-FG05-85ER250000 and DE-FG05-96ER40979.  Computations were performed
on the CM-2, the workstation cluster at SCRI, and the Xolas computing
cluster at MIT's Laboratory for Computing Science.
}

\vfill\eject

\begin{table}
\addtolength{\tabcolsep}{-1.0mm}
\begin{flushleft}
\begin{tabular}{|cc|cc|cc|}
\hline
 \multicolumn{2}{|c}{$c_{0}^{(6)} = 0.182031$} & 
 \multicolumn{2}{|c}{$c_{0}^{(8)} = 0.138827$} &
 \multicolumn{2}{|c|}{$c_{0}^{(10)} = 0.112209$}  \\
\hline
$P_6$ & $Q_6$ & $P_8$ & $Q_8$ & $P_{10}$ & $Q_{10}$ \\
2.05911386e-04 & 2.59134855e-05 & 8.76771341e-05 & 1.41842155e-05 & 4.66940445e-05 & 8.66147628e-06 \\ 
4.41581797e-03 & 1.09739974e-03 & 1.23623252e-03 & 3.61450316e-04 & 4.91389639e-04 & 1.63943675e-04 \\ 
3.77980526e-02 & 1.40770643e-02 & 9.11319209e-03 & 3.60062816e-03 & 3.13861891e-03 & 1.31054157e-03 \\ 
2.01546314e-01 & 9.03340757e-02 & 4.26239090e-02 & 2.05879480e-02 & 1.36974237e-02 & 6.82851948e-03 \\ 
1.00302812e+00 & 4.40335697e-01 & 1.54018532e-01 & 8.27820958e-02 & 4.57939021e-02 & 2.57134015e-02 \\ 
1.18344772e+01 & 2.67694284e+00 & 5.10199622e-01 & 2.80210832e-01 & 1.30067127e-01 & 7.83188602e-02 \\ 
 &  & 1.96142588e+00 & 9.59252529e-01 & 3.43184790e-01 & 2.12031104e-01 \\ 
 &  & 2.06291295e+01 & 4.86700520e+00 & 9.33072953e-01 & 5.59010141e-01 \\ 
 &  &  &  & 3.19171360e+00 & 1.64058378e+00 \\ 
 &  &  &  & 3.17877409e+01 & 7.65233478e+00 \\ 
\hline
\end{tabular}
\end{flushleft}
\vspace{5mm}
\caption{
The normalization and roots for the optimal rational function approximation to
$\epsilon(z)$ (Eq.~\ref{eq:ratio_factored}) over the interval 
$[0.005,1]$ for $N=6, 8$ and $10$.
}
\label{tab:remez}
\end{table}

\newpage

\begin{figure}
\epsfxsize=5in
\centerline{\epsffile{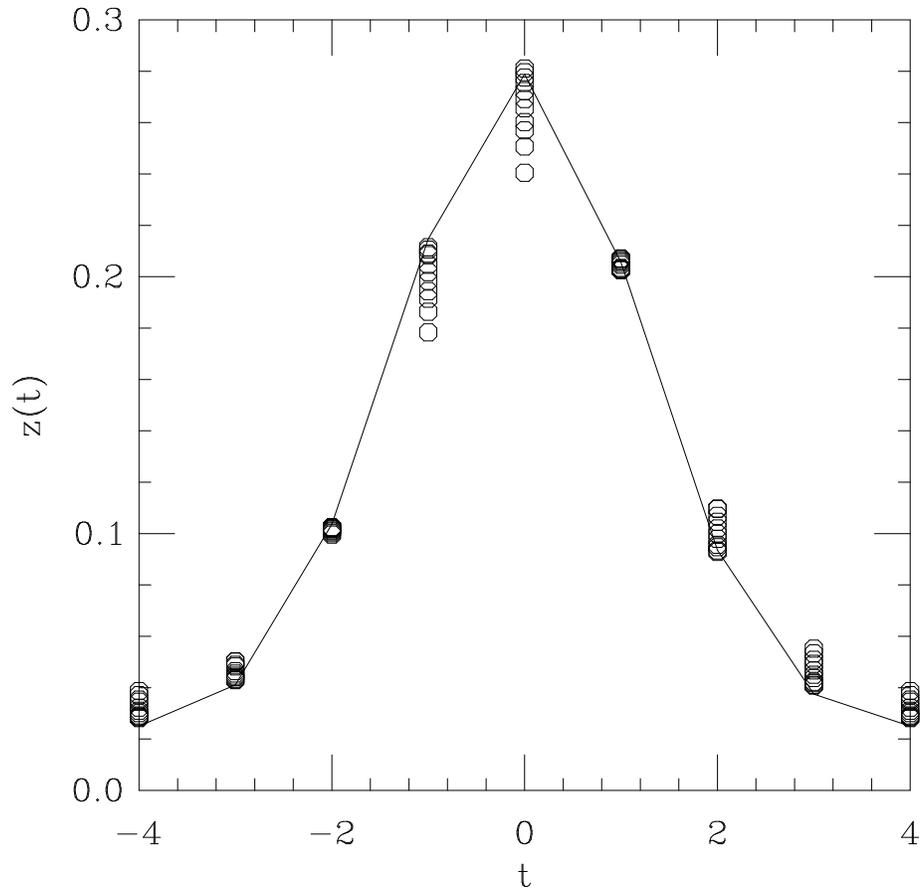}}
\caption{
The mode associated with the zero eigenvalues in
Fig.~\ref{fig:flow_cfg17} for $m > 0.8$ is plotted with octagon
symbols and compared with the zero mode of $\ham_w$ shown
as straight line segments.
}
\label{fig:zero_cfg17}
\end{figure}

\end{document}